\RequirePackage{ifpdf}
\documentclass[11pt,a4paper]{article}
\pdfoutput = 1
\usepackage{jcappub}

\let\a=\alpha
\let\e=\epsilon

\usepackage{color}
\usepackage[makeroom]{cancel}
\usepackage{multicol}

\usepackage{epsfig,bm,graphicx,caption}
\epsfclipon
\usepackage{amsmath,amssymb,amsbsy,amstext,amsthm}
\usepackage{subcaption}

\let\normalcolor\relax  

\newcommand{\roughly}[1]{\mathrel{\raise.3ex\hbox{$#1$\kern-0.85em
\lower1ex\hbox{$\sim$}}}}

\def\be{\begin{equation}}
\def\ee{\end{equation}}
\def\ba{\begin{eqnarray}}
\def\ea{\end{eqnarray}}

\usepackage{ulem}

\usepackage[T1]{fontenc} 

\title{\boldmath Theory of Cosmological Perturbations with Cuscuton}

\author[a]{Supranta S. Boruah}
\author[a]{ Hyung J. Kim} 
\author[a,b]{Ghazal Geshnizjani}

\affiliation[a]{Department of Applied Mathematics, University of Waterloo\\ Waterloo, Ontario, N2L 3G1, Canada}
\affiliation[b]{Perimeter Institute for Theoretical Physics\\ 31 Caroline St. N., Waterloo, ON, N2L 2Y5, Canada}

\emailAdd{ssarmabo@uwaterloo.ca}
\emailAdd{h268kim@uwaterloo.ca}
\emailAdd{ggeshniz@uwaterloo.ca}

\abstract{This paper presents the first derivation of the quadratic action for curvature perturbations, $\zeta$, within the framework of cuscuton gravity. We study the scalar cosmological perturbations sourced by a canonical single scalar field in the presence of cuscuton field. We identify $\zeta$ as comoving curvature with respect to the source field and we show that it retains its conservation characteristic on super horizon scales. The result provides an explicit proof that cuscuton modification of gravity around Friedmann-Lemaitre-Robertson-Walker (FLRW) metric is ghost free. We also investigate the potential development of other instabilities in cuscuton models. We find that in a large class of these models, there is no generic instability problem. However, depending on the details of slow-roll parameters, specific models may display gradient instabilities.}

\begin{document}
\maketitle
\flushbottom
\section{Introduction}\label{intro}
 
 Cuscuton gravity was originally proposed about a decade ago\cite{CuscTheory, CuscCosmo}, as an infrared modification of general relativity, with no additional degree of freedom. Interestingly, since then cuscuton has been rediscovered in other independent works to address different questions in early or late universe.  For instance, it rises \cite{CuscHorava1, CuscHorava2} in the low energy limit of the Horava gravity\cite{Horava}. It was also noticed to appear in new types of inflationary models \cite{Magnon}, in alternative models for inflation \cite{Bessada:2009ns}, new approaches to resolve cosmological constant problem \cite{D'Amico:2011jj, Afshordi:2014qaa}, etc. 
 In a recent work, a galileon generalisation of the cuscuton was realised to be one of the three generalisation of the galileon theories that do not form spherical caustics\cite{deRhamCuscuton}. 

 Cuscuton model can be simply formulated \cite{CuscTheory} by introducing a non-canonical scalar field to general relativity and requiring the field to be incompressible. One notices that in that limit the equation of motion of this field does not have any second order time derivatives. This means that the cuscuton field does not have its own dynamics, but rather modifies the dynamics of other dynamical fields. In other words, cuscuton acts as an auxiliary field and it does not introduce new degrees of freedom. It is also manifest through all the equations of motion that it only modifies general relativity on large scales.
 
Even though over the years many implications of cuscuton cosmology have been explored, until now the quadratic action for curvature perturbations were not explicitly obtained. Part of the reason for this is that cuscuton equation of motion is a constraint equation which introduces non-local operators in the action. To evade this problem, here we carry on our analysis in Fourier space. The other difficulty is that a priori, it is not clear what variable would be a best candidate for describing a conserved quantity $\zeta$, if it even exists in cuscuton models. In standard theory of cosmological perturbations, there are different ways one can define this quantity based on a particular gauge or matter components in the model \cite{Bardeen:1980kt, Bardeen:1983qw, MKBpert, Maldacena}. For instance, it can be defined as curvature perturbation with respect to the comoving gauge for one of the matter components or alternatively with respect to the total matter \cite{Bardeen:1980kt}. It can also be defined in terms of metric fluctuations and the over all equation of state in Newtonian/Longitudinal gauge \cite{MKBpert}. For single component models, these definitions either coincide or merge on super horizon scales. Furthermore, on these scales they all contain a conserved mode and a time dependent mode which often decays away. Now for cuscuton, even though it does not introduce any additional degree of freedom, it does resemble multi-field models. To be more explicit, its formulation starts by modifying the right hand side of Einstein equation. Naively, it seems that it contributes to energy density and momentum density. Therefore, the question is which of these definitions will be best suited for defining $\zeta$. In the end, it became evident to us that all these definitions generate a conserved mode at large scales and they merge on small scales. However, it turned out that if $\zeta$ is defined as comoving curvature perturbation with respect to only the source field, the computations are significantly simplified.   

Derivation of the action for $\zeta$ provides a rigorous proof that indeed cuscuton models do not have any ultra violet pathology. Furthermore, it provides fascinating new possibilities for beginning of our universe that could not be explored within the realm of general relativity. One example of that which we are working on is the possibility of a regular bounce initial condition with cuscuton. In general relativity that entails breaking energy conditions which lead to different types of instabilities. However, as our work shows cuscuton can evade breaking those conditions.

This paper is organized in the following way: In Section (\ref{cusc}), We review the background equations in a Friedmann-Lemaitre-Robertson-Walker (FLRW) universe in presence of cuscuton. We then outline the derivation of second order action for curvature perturbation in presence of cuscuton Field in section (\ref{secondorderaction}). In case readers are interested to repeat our analysis, a more detailed version of this derivation can be found in Appendix (\ref{details}). In section (\ref{ghost}), we discuss why the quadratic action for $\zeta$ implies that cuscuton theories are ghost free. We also study the conditions for appearance of other types of instabilities in cuscuton models and conservation of $\zeta$ in Infra Red (IR). Our concluding remarks are presented in (\ref{concl}). 

\section{Cuscuton}\label{cusc}

\subsection{Review of theory}\label{theory}

Consider a $P(X,\varphi)$ theory with, $X = \partial_{\mu}\varphi\partial^{\mu}\varphi$ on a FLRW cosmological background where metric is,
\begin{equation}\label{metric}
    ds^2 = a^2(\tau)(-d\tau^2+\delta_{ij}dx^idx^j).
\end{equation}
The field equation for such a theory is, 
\begin{equation}\label{PXeqn}
    (P_{,X}+2X P_{,XX})\varphi^{\prime\prime}+3\mathcal{H} P_{,X}\varphi^{\prime}+P_{,X\varphi}\varphi^{\prime 2}-\frac{1}{a^2}P_{,\varphi} = 0. 
\end{equation}
Considering the limit where the coefficient of the second derivative term vanishes, leads to
\begin{equation}\label{CuscDef}
    P_{,X}+2X P_{,XX} = 0.
\end{equation}
The unique theory where this condition is satisfied everywhere in phase space is given by the following action, 

\begin{equation}\label{CuscAction}
    S_{cusc} = \int d^4x\sqrt{-g}\big[\pm \mu^2 \sqrt{X} - V(\varphi)\big].   
\end{equation}
The field equation obtained from the action\eqref{CuscAction} is given by,
\begin{equation}\label{CuscEqn}
    (g_{\mu\nu}-\frac{\partial_{\mu}\varphi\partial_{\nu}\varphi}{X})D^{\mu}D^{\nu}\varphi\pm \frac{1}{\mu^2}\sqrt{X}V^{\prime}(\varphi) = 0, 
\end{equation}
where $D^\mu$ denotes the four dimensional covariant derivative. There are different ways that one can show this equation does not have any propagating modes \cite{CuscTheory}. This will be manifest in our analysis later in this article at linear order as well. However, the main argument holds at any order in perturbation theory.

\subsection{Background Cosmology}\label{background}

As mentioned above, cuscuton is a field with no dynamics and acts as a non-local modification to Einstein's gravity. Therefore, to produce dynamical cosmological solutions in a cuscuton scenario, there needs to be other sources with propagating degrees of freedom. In our work, we consider the scalar mode to be sources by a scalar field, $\pi$, with a canonical kinetic term and minimally coupled to cuscuton.
So we start with the action\footnote{We are setting the value of Planck mass to one.},
\begin{equation}\label{action}
    S = \int d^4x \sqrt{-g} \big[ \frac{1}{2}R - \frac{1}{2} D_\mu\pi D^\mu \pi-U(\pi)\pm \mu^2\sqrt{-D_\mu \varphi D^\mu \varphi}-V(\varphi) ]
\end{equation}
We now substitute the FLRW metric \eqref{metric} in this action and assuming homogeneity and isotropy, derive the the background equations 
\begin{subequations}
	\begin{equation}\label{Friedmann}
		3\mathcal{H}^2 = \frac{1}{2}\pi^{\prime 2}_{0} + V(\varphi_0) a^2 + U(\pi_0) a^2
	\end{equation}
	\begin{equation}\label{RCFriedmann}
		\mathcal{H}^2 -\mathcal{H}^{\prime}=  \frac{1}{2}\pi^{\prime 2}_{0} \pm \frac{\mu^2}{2} \mid \varphi^{\prime}_{0} \mid a.
	\end{equation}
\end{subequations} 
In our notations, $\mathcal{H} \equiv \frac{a^{\prime}}{a}$, where $^{\prime}$ denotes the derivative with respect to conformal time, $\frac{d}{d\tau}$. We are also denoting the background homogeneous values with a subscript, $_{0}$. If we introduce the following dimensionless quantities, 
\ba 
\alpha &\equiv& \frac{\pi^{\prime 2}_0}{2\mathcal{H}^2}\\
\epsilon &\equiv& \frac{\mathcal{H}^2-\mathcal{H}^{\prime}}{\mathcal{H}^2}, 
\ea 
then equation \eqref{RCFriedmann} can be written as 
\begin{equation}\label{defdelta}
	\sigma\equiv \epsilon-\alpha = \pm\frac{\mu^2}{2\mathcal{H}^2} \mid \varphi^{\prime}_{0} \mid a.
\end{equation}
 In standard single field models, the quantities $\epsilon$ and $\alpha$ coincide and in inflationary context, they are referred to as {\it the first slow roll} parameter.
Therefore, $\sigma$, indicates the deviations from standard GR due to cuscuton. Further more, its sign is dictated by the sign of $\mu^2$ taken in the action. Note that if we choose $+\mu^2$ in the action, $\epsilon$ will automatically be positive.

Next, we can obtain the equation of motion for cuscuton   
\begin{equation}\label{CuscBackgEqn1}
	\pm	3\mu^2 {\textrm{sign}(\varphi^{\prime}_0)} \mathcal{H} =-a V_{,\varphi}(\varphi_0), 
	\end{equation}
which as expected is only a constraint equation for $\mathcal{H}$. For a specific cuscuton potential, equation \eqref{CuscBackgEqn1} can be inverted to express $\varphi_0$ as a function of $\mathcal{H}$. Combining that with  \eqref{Friedmann}. we can then explicitly see that at background level, cuscuton simply modifies Friedmann equation 
\begin{equation}\label{ModifiedGravEqn}
		3\mathcal{H}^2 = \frac{1}{2}\pi^{\prime 2}_{0} + U(\pi_0) a^2 + V\bigg(V_{,\varphi}^{-1}\bigg(-\frac{\pm 3\mu^2 {\textrm{sign}(\varphi^{\prime}_0)} \mathcal{H}}{a}\bigg)\bigg) a^2.
\end{equation}
Therefore, the functional form of the potential $V(\varphi)$, dictates the form of modified Friedmann equation, \eqref{ModifiedGravEqn}. Also there is no consistent cuscuton FRW solution if cuscuton potential is set to zero. Equation \eqref{CuscBackgEqn1} also leads to 

	\begin{equation}\label{CuscBackgEqn2}
	V_{,\varphi\varphi}(\varphi_0)=\frac{3\mu^4}{2}\left(1+\frac{\alpha}{\sigma}\right)
	\end{equation}
which tells us if we choose $+\mu^2$ in the action ($\sigma>0$), then there is a lower bound on $V_{,\varphi\varphi}$. 
	
We end this section by including the equation of motion for the scalar field, $\pi$, 
\begin{equation}\label{ScalarBackgEqn}
		\pi^{\prime \prime}_0 + 2\mathcal{H}\pi^{\prime}_0 -a^2\frac{\partial U}{\partial\pi}\pi^{\prime}_0= 0.
	\end{equation}

\section{The quadratic action for curvature perturbations with Cuscuton}\label{secondorderaction}
This section presents our main result. Similar to the standard method of deriving the quadratic action we start with ADM formalism \cite{ADM}. ADM variables provide a convenient way for splitting the $3+1$ space-time into a space-like foliation and a time direction. In this approach, metric is written in terms of the lapse, $N$, shift, $N_{i}$ and the 3-dimensional metric $h_{ij}$ as,
\begin{equation}\label{ADMmetric}
    ds^2 = -N^2 d\tau^2 + h_{ij}(dx^i+N^id\tau)(dx^j+N^jd\tau).
\end{equation}
Rewriting the action \eqref{action} in terms of Eintein-Hilbert part, the scalar field, $\pi$ and Cuscuton part and then substituting for ADM variables we get 
\begin{equation}\label{ADMaction}
   S=S_{EH}+S_\pi+S_\varphi, 
\end{equation}
where
\begin{align}
S_{EH}&=\frac{1}{2}\int d\tau d^3x \sqrt{h}\bigg[NR^{(3)}+N^{-1}(E_{ij}E^{ij}-E^2)\bigg], \label{EHaction}\\
   S_{\pi}&=\frac{1}{2}\int d\tau d^3x \sqrt{h}\bigg[N^{-1}(\pi^{\prime}-N^i\partial_i \pi)^2 - Nh^{ij}\partial_i\pi \partial_j \pi-2N U(\pi)\bigg],\label{piaction}\\
   S_{\varphi}&=\frac{1}{2}\int d\tau d^3x \sqrt{h}\bigg[\pm 2\mu^2\sqrt{((\varphi^{\prime}-N^i\partial_i\varphi)^2-N^2h^{ij}\partial_i\varphi \partial_j \varphi)} - 2N V(\varphi)\bigg].\label{varphiaction}
\end{align}

$R^{(3)}$ represents the Ricci scalar of the spacial hyper-surfaces and $E_{ij}$ is defined as
\begin{equation}\label{ExtrCurv}
    E_{ij} = \frac{1}{2}h^{\prime}_{ij}-\frac{1}{2}(\nabla_{i}N_{j}+\nabla_{j}N_{i}).
\end{equation}
$\nabla$ represents the covariant derivative with respect to the spatial metric, $h_{ij}$, while $\partial$ denotes the partial derivative with respect to the comoving coordinates. Variation of action \eqref{ADMaction} with respect to lapse and shift leads to momentum and hamiltonian constraints,
\begin{subequations}
    \begin{equation}\label{MomCons}
        \nabla_{i}(N^{-1}(E^i_j-\delta^i_j E)) = q_{,i}
    \end{equation}
    \begin{equation}\label{HamCons}
        R^{(3)}+N^{-2}(E^2-E^{ij}E_{ij}) = 2\rho. 
    \end{equation}
\end{subequations}
Here $q_{,i}$ is the Momentum density and the $\rho$ is the total energy density, including cuscuton contributions.
We now proceed to perform perturbative analysis around FLRW background. 
There are two gauge degrees of freedom associated with the scalar perturbations. We can remove one of them by choosing uniform field gauge with respect to $\pi$ field
\begin{equation}\label{GaugeChoice}
    \delta\pi = 0.
\end{equation}
The other one can be fixed by setting the off-diagonal components of the spatial metric to zero\footnote{We are using similar convention and notations as \cite{Maldacena}.}
\begin{equation}
    \quad h_{ij} = a^2(1+2\zeta)\delta_{ij}~.
\end{equation}
In literature, $\zeta$ is often referred to as comoving curvature perturbation. Note that comoving here refers only with respect to $\pi$ field. As we will see this particular choice produces a viable conserved quantity and makes the computations considerably simpler. 
The scalar contributions to laps and shift function in the metric can be written as,
\begin{equation}\label{ScalarPert}
    N_{i} = \nabla_{i}\psi, \quad  N = a~(1+N_1)~.
\end{equation}
Finally, we denote the perturbations associated with the the cuscuton field by $\delta\varphi$. Writing the momentum constraint \eqref{MomCons} and the hamiltonian constraint \eqref{HamCons} to linear order in perturbations yields, 

\begin{subequations}
    \begin{equation}\label{MomConsGauge}
        N_{1} =  \frac{\zeta^{\prime}}{\mathcal{H}}\pm \frac{1}{2}\mu^2a~ \textrm{sign}(\varphi_{0}^{\prime})\frac{\delta \varphi}{\mathcal{H}}
    \end{equation}
    \begin{equation}\label{HamConsGauge}
        \nabla^2\psi = -\frac{1}{\mathcal{H}}\nabla^2\zeta +\frac{\pi_0^{\prime 2}}{2\mathcal{H}}N_1~.
    \end{equation}
\end{subequations}
The next step is to perturb action \eqref{ADMaction} to second order in perturbative variables, $N_1$, $\psi$, $\zeta$ and $\delta\varphi$. This calculations is tedious and readers can refer to Appendix(\ref{details}) for the details. We then remove $N_1$ and $\psi$ using the constraint equations \eqref{MomConsGauge} and \eqref{HamConsGauge}. The result for the second order action after taking into account the background equations, is 
\begin{equation}\label{ActionZeta}
	S^{(2)} = \int d\tau d^3x \; a^2 \left[ \alpha\zeta^{\prime 2}-\epsilon(\partial \zeta)^2+\sigma \left ( \frac{\mathcal{H} \delta\varphi}{\varphi'_0}\right )\big(\alpha \mathcal{H}\zeta^{\prime}-\partial^2\zeta\big)\right]~.
\end{equation}
 In the $\sigma\rightarrow 0$ limit that contributions from cuscuton vanish, action \eqref{ActionZeta}, simplifies to the standard quadratic action for curvature perturbations. 
\begin{equation}\label{ActionZetaST}
	S^{(2)} = \int d\tau d^3x \; a^2 \alpha\left[\zeta^{\prime 2}-(\partial \zeta)^2\right]
\end{equation}

As we pointed out before, the field equation for cuscuton \eqref{CuscEqn} provides another constraint equation. At linear order this equation reduces to 
\begin{equation}\label{CuscEqnGauge}
\nabla^2\delta\varphi- \mathcal{H}^2 \alpha \big [3+\alpha-\epsilon \big ] \delta\varphi= \frac{\varphi^{\prime}_0}{\mathcal{H}}\left [\nabla^2\zeta-\alpha\mathcal{H}\zeta^{\prime}\right].
\end{equation}

In order to eliminates $\delta\varphi$ from action \eqref{ActionZeta}, we need to invert the above equation. However, since this involves inverting derivative operators, we continue our derivation in fourier space. This allows us to substitute for $\delta \varphi_k$ in terms of $\zeta_k$ and $\zeta_k^{\prime}$ using this formula  

\begin{equation}\label{CuscEqnFourier}
    \delta \varphi_k = \frac{\varphi^{\prime}_{0}}{\mathcal{H}}\frac{\textbf{k}^2\zeta_k+\alpha  \mathcal{H} \zeta^{\prime}_k}{\big[\textbf{k}^2+(3+\alpha -\epsilon)\alpha \mathcal{H}^2 \big]}.
\end{equation}

After Fourier transforming action \eqref{ActionZeta}, substituting for $\delta \varphi_k$ and some algebraic calculations, we finally arrive at, 
\begin{equation}\label{2ndorderaction}
    S^{(2)} = \int d^4x ~ z^2 \bigg[ \zeta^{\prime 2}_k-c_s^2~\textbf{k}^2 \zeta^2_k\bigg]. 
\end{equation}
$z(\textbf{k},\tau)$ and $c_s(\textbf{k}, \tau)$ in above action are both time and scale dependent functions given by
\ba
z^2&\equiv &a^2 \alpha \bigg(\frac{\textbf{k}^2+3\alpha \mathcal{H}^2}{\textbf{k}^2+\alpha \mathcal{H}^2 (3-\sigma)}\bigg) \label{Z2}\\
c_s^2 & \equiv & \frac{\textbf{k}^4+\textbf{k}^2\mathcal{H}^2\mathcal{B}_1+\mathcal{H}^4\mathcal{B}_2}{\textbf{k}^4+\textbf{k}^2\mathcal{H}^2\mathcal{A}_1+\mathcal{H}^4\mathcal{A}_2},
\ea
and we have introduced the following additional notations,\footnote{A detailed calculation is presentation in Appendix A.}
\begin{align}\label{defn1}
\eta &\equiv \frac{\epsilon^{\prime}}{\mathcal{H} \epsilon} \\
    \beta &\equiv \frac{\alpha^{\prime}}{\mathcal{H} \alpha}\\
    \mathcal{A}_1 &\equiv 6\alpha-\a \sigma\\
    \mathcal{A}_2 &\equiv 9\alpha^2-3\a^2\sigma\\
	\mathcal{B}_1 &\equiv \mathcal{A}_1+\sigma (6+\eta+\beta-2\e)+\alpha(\eta-\beta)\\
    \mathcal{B}_2 &\equiv \mathcal{A}_2+\sigma\alpha(12-4\sigma+3\eta)+3\alpha^2(\eta-\beta).
    \end{align}
It is also evident here that in the $\sigma\rightarrow 0$ limit, we get back the standard single scalar field result of $c_s^2\sim 1$ and $z^2\sim a^2\a$. 


\section{Ghosts, instabilities and conservation of $\zeta$} \label{ghost}

It is manifest from action \eqref{2ndorderaction}, that cuscuton is ghost free around FLRW background. As we had expected the leading $\textbf{k}$ terms in the action, do not have cuscuton dependence. Therefore, in UV limit ($\textbf{k} \rightarrow \infty $) we get the standard single scalar field result of $z^2\sim a^2\a >0 $. This implies that the theory is ghost free regardless of sign or value of $\e$, or which sign for $\mu^2$ is taken in the action. In fact, one generic feature is that for $-\mu^2$ in the action, since $\sigma$ is automatically negative, $z^2$ always remains positive regardless of scale. On the other hand if we pick the $+\mu^2$ factor in the original  cuscuton action \eqref{CuscAction}, then $\sigma>0$. In this case, one may ask what happens in a region of parameter space with $\sigma\geq 3$. In other words, is there a pathology associated to $z^2$ diverging or becoming negative. Note that producing such a model would require engineering peculiar potentials and tuning of $\mu^2$ which seems very contrived. Nevertheless, that would not indicate a ghost in the theory. The notion of ghost is only a meaningful statement in the UV limit and as we have pointed earlier, that limit is always fine. When we deviate from the limit of flat background or time independent actions, energy conservation and plain wave description of modes breaks down. One may still evaluate the Hamiltonian and it can be negative but that doesn't necessarily tell us if there is an instability in the system or not. In fact even in standard inflationary models, Hamiltonian becomes negative on super horizon scales and resembles excited states with negative energy but theory is still healthy\cite{mukhanov2007introduction}. 

A theory might be ghost free but still suffer from other types of instabilities such as gradient instability. However, as long as instabilities are not in ultra violet, they are only indicative of a growing solution that can be circumvented by tuning the parameters of the model. Whether, a particular cuscuton scenario exhibits such an instability in a specific region of phase space or not, will depend on details of the model. To elaborate  that let us write down the equation of motion for $\zeta_k$ derived from action \eqref{2ndorderaction} 
\begin{equation}
	\zeta^{\prime\prime}_k+\bigg(2+\beta+\frac{\mathcal{C}_1\mathcal{H}^2k^2+\mathcal{C}_2\mathcal{H}^4}{\textbf{k}^4+\textbf{k}^2\mathcal{H}^2\mathcal{A}_1+\mathcal{H}^4\mathcal{A}_2} \bigg) \mathcal{H}\zeta^{\prime}_k+c_s^2k^2 \zeta_{k}=0
\end{equation}
where,
\begin{align}
	\mathcal{C}_1 &= (\beta+2\a-2\a^2-2\a\sigma)\sigma+3\alpha^2(\eta-\beta)\\
    \mathcal{C}_2 &= 3\a^2(\eta-\beta).
\end{align}
As we see there are quite a few parameters that can determine the sign and behaviour of $c_s^2$ and coefficients of $\zeta_k'$. While we can not make conclusive statement for every cuscuton scenario, we comment on some generic features.  
First, in UV limit all the cuscuton contributions go away and $c_s^2\rightarrow 1$. Therefore, there is no gradient instability in that limit. 

Second, the non trivial denominator shared in one of the coefficients of $\zeta_k'$ and $c_s^2$ can be factored as 
\be
\textbf{k}^4+\textbf{k}^2\mathcal{H}^2\mathcal{A}_1+\mathcal{H}^4\mathcal{A}_2=(\textbf{k}^2+3\alpha \mathcal{H}^2)(\textbf{k}^2+\alpha \mathcal{H}^2 (3-\sigma)).
\ee
Therefore, for $-\mu^2$ in the action or $+\mu^2$ with $\sigma< 3$, the equation of motion for $\zeta_k$ is not singular. Models with $+\mu^2$ and $\sigma\geq 3$ can allow for poles which make the equation of motion for $\zeta_k$ singular. Singular ODEs are not necessarily catastrophic and they may be treatable.
In fact as we mentioned before for $+\mu^2$, Eq. \ref{defdelta}, dictates that $\e>0$ at all times. Therefore, an expanding universe can not go through a bounce. It turns out for $+\mu^2$ and $\e>0$, one can write the equation of motion for $\Phi$ potential in longitudinal gauge and there the equation is not even singular \cite{CuscCosmo}. 

Next, we check the behaviour of $\zeta_k$ in IR to see if it is conserved or not. 
 In $k\rightarrow 0$ limit as long as $\sigma\neq 3$,  $z^2 c_s^2$ remains finite and equation of motion can be estimated as 
\be
\frac{d}{d\tau} z^2\zeta_k'\approx 0. 
\ee
Similar to the standard case, the solutions to this equation include a desirable constant mode for $\zeta_k$ as well as a time dependent mode that goes as $\int d\tau /z^2$. One can investigate under what conditions this mode decays away or grows outside horizon. Substituting $z^2$ from Eq. \ref{Z2}, taking the  IR limit and rewriting the time dependence of this mode in terms of e-folding number, $N\equiv \ln{a} $, we find
\be 
\zeta_{IR}^{(time)}\propto \int \frac{d\tau}{z^2} \bigg \lvert_{IR}\approx\int \left (\frac{1-\frac{\sigma}{3}}{\a}\right )\left(\frac{dN}{e^{3N-\int\e d\tilde{N}}}\right).
\ee
Therefore, generically in an expanding universe ($N$ is increasing in time), $\e<3$ can lead to a decaying mode outside the horizon but $\e\geq 3$ can produce a growing mode. 
On the other hand, in a contracting model since $N$ is decreasing, we expect the reverse. Of course, this is no different from ordinary cosmological perturbation theory, except that here a cuscuton model may compensate for these effects by having the time variation of $\sigma$ cancel the exponential term in the integral. It is also interesting to note that $\sigma\geq 3$ models which as we said can only be realized in $+\mu^2$ actions and expanding scenarios lead to $\e= \a+\sigma>3$.

Last, let us also comment on how are definition of $\zeta$ differs from other definitions in literature. For example, in single field models, sometimes a conserved parameter $\zeta_s$ is defined in longitudinal/Newtonian gauge\footnote{In this gauge, shift function, $N_i$ is set to zero and $h_{ij}=a^2(1+2\Phi)\delta_{ij}$ which implies $N_1=-\Phi$.} as 
\begin{equation}
\zeta_s=\Phi+{\Phi'+\mathcal{H}\Phi \over \epsilon \mathcal{H}}. 
\end{equation}
A time transformation $t\rightarrow t-\psi$ shows that $\zeta_s$ is related to our choice of $\zeta$  in the following way
\begin{equation}
    \zeta_s=\zeta-{\sigma\mathcal{H} \delta\varphi\over \varphi_0'}. 
\end{equation}
Substituting for $\delta\varphi$ from \eqref{CuscEqnFourier}, we can obtain an explicit relation for $\zeta$ to the Newtonian potential $\Phi$ in Fourier space
\begin{equation}
\zeta_k=\Phi_k+{\Phi_k'+\mathcal{H}\Phi_k \over \epsilon \mathcal{H}} \left[1- {3\mathcal{H}^2\sigma\over \textbf{k}^2+3\mathcal{H}^2\epsilon}\right] . 
\end{equation}
As we see these equations show that in  $\sigma\rightarrow 0 $ and UV limit, these two definitions merge. In IR limit we get
\begin{equation}
    \zeta_s\simeq\zeta+{\sigma\over \mathcal{H}(3-\sigma)}\zeta', 
\end{equation}
which implies if $\zeta$ is conserved $\zeta_s$ will be conserved too. We can also perform the time transformation $t\rightarrow t+{\sigma\over\e}{\delta\varphi\over \varphi_0'}$ to go the comoving gauge with respect to the total momentum of both cuscuton and $\pi$ field ($T^0_i=0$). In that case the comoving curvature perturbation is 
\begin{equation}
    \zeta_t=\zeta+{\sigma\mathcal{H} \delta\varphi\over \epsilon \varphi_0'},  
\end{equation}
which leads to similar results in different limits. Therefore, from physical point of view there does not seem to be any advantage in choosing one definition over the other as long as $\zeta$ does not have a growing mode outside the horizon. However, from computation point, we found that derivation of equations and action were considerable simpler when we used the comoving gauge with respect to the source field.  

We end this discussion by emphasising again that similar to usual model buildings in GR scenarios, the question of instabilities will depend a lot on details of the potentials. If anything, with cuscuton there is more room to evade these problems.


\section{Conclusion}\label{concl}
The main goal of this paper was to obtain the quadratic action for comoving curvature perturbations, $\zeta$, in cuscuton models. We started from an action that included the standard Hilbert-Einstein term, a canonical scalar field and a cuscuton field. We then used ADM formalism and the uniform field gauge with respect to the scalar field to obtain the quadratic action for  $\zeta$. In order to eliminate the cuscuton dependence from this action we had to invert the cuscuton constraint equation. Therefore, we carried on the derivation in Fourier Space. As we expected our final action \eqref{2ndorderaction} had a complicated form but it explicitly shows that cuscuton models are ghost free and have no instabilities in UV limit. Basically in UV limit, the action becomes the standard quadratic action for a scalar field, minimally coupled to gravity. Upon further investigation of equation of motion for $\zeta_k$ in section \ref{ghost}, we also saw that there are no out of ordinary instabilities on non-UV scales either. This analysis shows that depending on the details of a particular cuscuton model and the potential of the scalar field, some corners of parameter space may lead to growing modes. Interestingly, it seems if we choose a $-\mu^2$ for cuscuton kinetic term in the action, there is more flexibilities in engineering different background evolutions and less chance of developing instabilities. That will be very useful in engineering bounce scenarios. In order to get a bounce one has to choose $-\mu^2$ in the action and make the parameter $\e$ become negative. However, with cuscuton that does not lead to ghosts since the source field does not violate null energy condition. That is the subject of our next upcoming paper. We also showed that our choice of $\zeta$ was consistent with producing a conserved mode on super horizon scales. However, we noticed that other common definitions of $\zeta$ while are different physical quantities, they also produce a conserved mode and all of these definitions merge on small scales. From computation point, we found that derivation of equations and action were considerable simpler when we picked $\zeta$ as the comoving curvature perturbation with respect to the source field.  
\acknowledgments
This research project was supported by the Discovery Grant from Natural Science
and Engineering Research Council of Canada.  GG is also supported in part by Perimeter Institute.
Research at Perimeter Institute is supported by the Government of Canada through
the Department of Innovation, Science and Economic Development Canada and by
the Province of Ontario through the Ministry of Research, Innovation and Science.

\appendix 
\section{More details about the calculation of second order action}\label{details}

In this appendix we present some of the intermediate steps of our derivation in section \ref{secondorderaction}. 
After fixing the gauge, we perturb different parts of the action \eqref{ADMaction} to second order in perturbative variables, $N_1$, $\psi$, $\zeta$ and $\delta\varphi$. We then remove $N_1$ and $\psi$ using the constraint equations \eqref{MomConsGauge} and \eqref{HamConsGauge}. The result after taking into account the background equations,  \eqref{Friedmann}, \eqref{RCFriedmann} and \eqref{CuscBackgEqn1} is

\begin{align}
	&S_{EH}^{(2)} &= \int d\tau d^3x \; & a^2 \left\{ \left[\frac{3\epsilon}{2}-9\right](\zeta\mathcal{H})^2-\epsilon(\partial\zeta)^2+\mu^2a\delta\varphi\left[\alpha\zeta^{\prime}-\frac{9}{2}\zeta\mathcal{H}\right]\right. \nonumber \\
	& & &\left. -(\mu^2a\delta\varphi)^2\left[\frac{\alpha}{2}+\frac{3}{4}\right] \right\} \label{ActionZetaG}\\ 
	\label{ActionZetaC}
	&S_{\varphi}^{(2)} &= \int d\tau d^3x \; & a^2  \left\{\left[6(\alpha-\epsilon)+\left[\frac{3Va^2}{\mathcal{H}^2}\right]\left(1+\frac{\epsilon}{2}\right)\right](\zeta\mathcal{H})^2+\mu^2a\delta\varphi\left[\frac{3a^2V}{2\mathcal{H}^2}(\zeta\mathcal{H}) \right.\right. \nonumber\\
	& & &\left.\left.-\frac{1}{2}\alpha\zeta^{\prime}+\frac{\partial^2\zeta}{2\mathcal{H}}\right] +(\mu^2a\delta\varphi)^2\left(\frac{\alpha}{4}+\frac{3}{4}\right) \right\}\\
\label{ActionZetaM}
	&S_{\pi}^{(2)} &= \int d\tau d^3x \; &a^2 \left\{\alpha \zeta^{\prime2}+\left[\left(9-6\alpha+\frac{9\epsilon}{2}\right)-\frac{3Va^2}{\mathcal{H}^2}\left(1+\frac{\epsilon}{2}\right)\right](\zeta\mathcal{H})^2\right. \nonumber \\
	& & & \left.+\mu^2a\delta\varphi\left[\left(\frac{9}{2}-\frac{3a^2V}{2\mathcal{H}^2}\right)\zeta\mathcal{H}-\alpha\zeta^{\prime}\right]
	+\frac{1}{4}(\mu^2a\delta\varphi)^2\right\} 
\end{align}
Combining these expressions we obtain the expression \ref{ActionZeta}, 

\begin{equation}
	S^{(2)} = \int d\tau d^3x \; a^2 \left[ \alpha\zeta^{\prime 2}-\epsilon(\partial \zeta)^2+\sigma \left ( \frac{\mathcal{H} \delta\varphi}{\varphi'_0}\right )\big(\alpha \mathcal{H}\zeta^{\prime}-\partial^2\zeta\big)\right]~.
\end{equation}
We then proceeded to eliminate $\delta\varphi$ from above action while continuing our derivation in Fourier space. We substituted for $\delta \varphi_k$ in terms of $\zeta_k$ and $\zeta_k^{\prime}$ using equation \eqref{CuscEqnFourier} and obtained

\begin{align}
	S^{(2)} &= \int d^4x a^2\big[ \alpha\zeta_k^{\prime 2}-\epsilon(\partial \zeta_k)^2+\sigma\frac{(\textbf{k}^2 \zeta_{k}+\alpha \mathcal{H}\zeta^{\prime}_{k})^2}{\textbf{k}^2+\alpha(3-\sigma)\mathcal{H}^2}\big]\nonumber \\
    &= \int d^4x a^2 \alpha\bigg[\bigg(\frac{\textbf{k}^2+3\alpha\mathcal{H}^2}{\textbf{k}^2+\alpha(3-\sigma)\mathcal{H}^2}\bigg)\zeta^{\prime 2}_{k}- \textbf{k}^2\bigg(\frac{\textbf{k}^2+(3-\sigma)\epsilon\mathcal{H}^2}{\textbf{k}^2+(3-\sigma)\alpha\mathcal{H}^2}\bigg)\zeta^2_{k}\nonumber \\
    &+\bigg(\frac{2\textbf{k}^2 \sigma\mathcal{H}}{\textbf{k}^2+(3-\sigma)\alpha\mathcal{H}^2}\bigg)\zeta_{k}\zeta^{\prime}_{k}\bigg] \nonumber\\
    &= \int d^4x a^2\bigg[ \alpha \bigg(\frac{\textbf{k}^2+3\alpha\mathcal{H}^2}{\textbf{k}^2+\alpha(3-\sigma)\mathcal{H}^2}\bigg)\zeta^{\prime 2}_{k}-\alpha \textbf{k}^2\bigg(\frac{\textbf{k}^2+(3-\sigma)\epsilon\mathcal{H}^2}{\textbf{k}^2+(3-\sigma)\alpha\mathcal{H}^2}\bigg)\zeta^2_{k} \nonumber \\
    &+\frac{\textbf{k}^2}{a^2}\bigg(\frac{\alpha a^2\sigma\mathcal{H}}{\textbf{k}^2+\alpha(3-\sigma)\mathcal{H}^2}\bigg)^{\prime}\zeta_{k}^2\bigg], 
\end{align}
where in the last step we applied integration by parts. After performing the algebraic evaluation of these term and introducing the second slow roll parameters $\eta$ and $\beta$ as
\begin{align}
\eta &\equiv \frac{\epsilon^{\prime}}{\mathcal{H} \epsilon} \\
    \beta &\equiv \frac{\alpha^{\prime}}{\mathcal{H} \alpha}
    \end{align} 
    we finally get 
\begin{equation}
    S^{(2)} = \int d^4x  a^2 \alpha\bigg[ \bigg(\frac{\textbf{k}^2+3\alpha \mathcal{H}^2}{\textbf{k}^2+\alpha \mathcal{H}^2 (3-\sigma)}\bigg)\zeta^{\prime 2}_k-\bigg( \frac{\textbf{k}^4+\textbf{k}^2\mathcal{H}^2\mathcal{B}_1+\mathcal{H}^4\mathcal{B}_2}{[\textbf{k}^2+\alpha \mathcal{H}^2 (3-\sigma)]^2}\bigg)\textbf{k}^2 \zeta^2_k\bigg],
\end{equation}
where $\mathcal{B}_1$ and $\mathcal{B}_2$ are given by relations 
 \begin{align}
	\mathcal{B}_1 &= 6\alpha+\sigma(\eta+6+\beta-2\sigma-3\alpha)+\alpha(\eta-\beta)\\
    \mathcal{B}_2 &= 9\alpha^2-\sigma\alpha(3\alpha+4\sigma-3(4+\eta))+3\alpha^2(\eta-\beta).
\end{align}
After identifying the coefficient of the kinetic and the gradient terms as $z^2$ and $z^2 c_s^2$, the final action can be presented as \eqref{2ndorderaction}.

\bibliographystyle{JHEP}

\bibliography{draft}

\end{document}